\documentclass{article}
\usepackage[top=2cm,bottom=2cm,right=2cm,left=2cm]{geometry}

\usepackage[utf8]{inputenc}
\usepackage{graphicx} 
\usepackage{subfig}
\usepackage[usenames, dvipsnames]{color}  
\usepackage[usenames, dvipsnames]{xcolor}  
\usepackage{soul}
\usepackage[normalem]{ulem}
\usepackage{siunitx}
\usepackage{hyperref}
\usepackage{amsmath}
\usepackage{amssymb}
\usepackage{multirow}
\usepackage{multicol}
\usepackage{printlen}
\usepackage{makecell}
\usepackage{xspace}
\usepackage[skins]{tcolorbox}
\tcbuselibrary{breakable}
\usepackage[switch]{lineno}
\usepackage{sectsty}
\usepackage{authblk}
\usepackage[toc,page]{appendix}
\usepackage{csquotes}
\usepackage{cite}
\usepackage{cuted}

\newif\ifprintcomments

\printcommentstrue

\newcommand{\hiddencomment}[1]{}

\tcbset{commonstyle/.style={boxrule=0pt,sharp corners,enhanced jigsaw,nobeforeafter,boxsep=0pt,left=\fboxsep,right=\fboxsep,enforce breakable}}
\newtcolorbox{mycolorbox}[1][]{commonstyle,#1}

\colorlet{soulSkyBlue}{SkyBlue!10}
\colorlet{soulRubineRed}{RubineRed!10}

\newcommand{\gracedb}{GraceDB\xspace}
\newcommand{\far}{\ensuremath{\mathrm{FAR}}\xspace}

\newcommand{\snropt}{SNR optimizer\xspace}
\newcommand{\pastro}{\ensuremath{p_\text{astro}}\xspace}
\newcommand{\pbns}{\ensuremath{p_\text{BNS}}\xspace}
\newcommand{\pbbh}{\ensuremath{p_\text{BBH}}\xspace}
\newcommand{\pnsbh}{\ensuremath{p_\text{NSBH}}\xspace}
\newcommand{\pterr}{\ensuremath{p_\text{terr}}\xspace}

\newcommand{\Msun}{\ensuremath{\mathrm{M_{\odot}}}\xspace}
\newcommand{\hertz}{\ensuremath{\mathrm{Hz}}\xspace}
\newcommand{\peryr}{\ensuremath{\mathrm{year^{-1}}}\xspace}
\newcommand{\hasssm}{\ensuremath{\text{HasSSM}}\xspace}
\newcommand{\hasns}{\ensuremath{\text{HasNS}}\xspace}
\newcommand{\hasmg}{\ensuremath{\text{HasMassGap}}\xspace}
\newcommand{\hasrem}{\ensuremath{\text{HasRemnant}}\xspace}

\newcommand{\mchirp}{\ensuremath{m_{\text{chirp}}}\xspace}

\newcommand{\mtot}{\ensuremath{m_{\text{tot}}}\xspace}

\newcommand{\spineff}{\ensuremath{\chi_\mathrm{eff}}\xspace}
\newcommand{\snr}{\ensuremath{\mathrm{SNR}}\xspace}

\newcommand{\soft}[1]{\textsc{#1}}
\newcommand{\STTFOUR}{\soft{SpinTaylorT4}\xspace}
\newcommand{\SEOBNRFOUROPT}{\soft{SEOBNRv4\_opt}\xspace}

\newcommand{\popone}{\textit{pop-1}\xspace}
\newcommand{\poptwo}{\textit{pop-2}\xspace}
\newcommand{\OFouraevents}{\textit{O4a-events}\xspace}

\newcommand{\tripledash}{\xspace---\xspace}

\title{Enhancing online estimation of CBC parameters with the low-latency MBTA analysis}

\author[1]{Florian Aubin\footnote{Corresponding author: \href{mailto:florian.aubin@iphc.cnrs.fr}{florian.aubin@iphc.cnrs.fr}}}
\author[2]{Inès Bentara}
\author[3]{Damir Buskulic}
\author[4,5]{Gianluca M Guidi}
\author[6]{Vincent Juste}
\author[2]{Morgan Lethuillier}
\author[3]{Frédérique Marion}
\author[4,5]{Lorenzo Mobilia}
\author[1]{Benoît Mours}
\author[2]{Amazigh Ouzriat}
\author[1]{Thomas Sainrat}
\author[2]{Viola Sordini}

\affil[1]{Université de Strasbourg, CNRS, IPHC UMR 7178, F-67000 Strasbourg, France}
\affil[2]{Institut de Physique des 2 Infinis de Lyon (IP2I) - UMR 5822, Université de Lyon, Université Claude Bernard, CNRS, F-69622 Villeurbanne, France}
\affil[3]{Univ. Savoie Mont Blanc, CNRS, Laboratoire d’Annecy de Physique des Particules - IN2P3, F-74940 Annecy, France}
\affil[4]{Università degli Studi di Urbino ’Carlo Bo’, I-61029 Urbino, Italy}
\affil[5]{INFN Sezione di Firenze, Sesto Fiorentino, I-50019 , Firenze, Italy}
\affil[6]{Service de Physique Théorique, Université Libre de Bruxelles (ULB), Boulevard du Triomphe, CP225, B-1050 Brussels, Belgium}

\date{}

\begin{document}

\maketitle

\begin{abstract}

In this paper, we describe the procedure implemented in the Multi-Band Template Analysis (MBTA) search pipeline to produce online posterior distributions of compact binary coalescence (CBC) gravitational-wave parameters.
This procedure relies on an \snropt{} technique, which consists of filtering dense local template banks. 
We present how these banks are constructed using information from the initial detection and detail how the results of the filtering are used to estimate source parameters and provide posterior distributions. 
We demonstrate the performance of our procedure on simulations and compare our source parameter estimates with the results from the first part of the fourth observing run (O4a) recently released by the LIGO-Virgo-KAGRA (LVK) collaboration.

\end{abstract}

\twocolumn

\section{Introduction} \label{sec:01}
The LIGO-Virgo-KAGRA (LVK) \cite{LIGO, Virgo, KAGRA} collaboration has been detecting gravitational wave (GW) signals from compact binary coalescences (CBCs) since 2015 \cite{GW150914}. 
Prior to compiling catalogs of sources \cite{GWTC4_results}, the collaboration performs low-latency searches \cite{PyCBC1, PyCBC2, GstLAL1, GstLAL2, GstLAL3, GstLAL4, SPIIR, MBTAO4, cWB1, cWB2, oLib, MLy} to rapidly identify candidates and release relevant information to the global astronomical community, enabling searches for electromagnetic (EM) or neutrino counterparts \cite{Mutlimessenger}.

During the second observing run (O2\footnote{Observing runs since 2015 are denoted O1, O2, O3 (divided into O3a and O3b) and O4 (divided into O4a, O4b and O4c) \cite{GWTC4_intro}.}), GW170817 was the first binary neutron star (BNS) merger to be observed \cite{GW170817}.
The sky localization derived from the GW signal, combined with the coincident short Gamma-ray burst \cite{GW170817_GRB}, enabled the detection of the source across multiple EM bands \cite{GW170817_Xray, Soares_Santos_2017, Cowperthwaite_2017, Nicholl_2017}, marking a milestone in multi-messenger astronomy.
This single event has led to numerous scientific breakthroughs \cite{GW170817_GRB, GW170817Strontium, GW170817_EOS, GW170817_H0}.

Current observing runs reveal that most detections involve binary black holes (BBHs), whereas mergers involving neutron stars remain rare. 
With the foreseen improvements in detector sensitivity \cite{LVK_prospects, ET_CE_Science}, the number of GW detections is expected to increase significantly. 
With limited capacity for electromagnetic observations, astronomers must carefully select which events to follow-up.
Confusion between low-mass black holes and neutron stars can lead to wasted effort, making it essential for low-latency CBC searches to rapidly provide reliable source characterizations that help the community prioritize the most promising candidates for multi-messenger observations.

This information is currently released in the form of sky maps, which provide the probable position of the source, along with source classification and properties \cite{UserGuide}.
The size of the sky maps mainly depends on the number of observing detectors, with the localization constrained by the arrival times, phases, and amplitudes measured across the network.
Low-latency source classification is obtained by modeling the pipeline’s response to astrophysical populations \cite{Kapadia_2020, Andres_2022, Villa_Ortega_2022, Farr_2015}. 
Refined source classifications are later released after using Bayesian parameter estimation (PE) algorithms \cite{Veitch_2015}, typically within tens of minutes by the RapidPE–RIFT pipeline \cite{Pankow_2015, rose2022}, and up to a few days by the more in-depth Bilby analysis \cite{Ashton_2019}, which also updates the source localization.
To maximize the detection of EM counterparts in future CBC observations, the main challenge is to produce reliable estimates of these and other observables as quickly as possible.
\\

In this paper, we introduce a solution to this problem implemented as a follow-up to the MBTA search \cite{MBTAO4}. 
Like other modeled GW searches conducted within the LVK \cite{GWTC4_methods}, MBTA analyzes data from each detector using a matched-filtering technique, with the specific approach of splitting the process across multiple frequency bands (typically two) to reduce computational costs.
The strain data time series are correlated with pre-generated CBC waveforms, commonly called \textit{templates}. 
These templates are elements of template banks, which are designed to provide a discrete sampling of the targeted parameter space.
By applying a threshold to the local maxima in signal-to-noise ratio (\snr{}) time series obtained by filtering the data against all the templates, the search produces a set of \textit{triggers}.
After undergoing a series of transient noise rejection procedures, sufficiently significant triggers are released to the \gracedb platform \cite{GraceDB, GraceDB_note}.

For each candidate, the pipeline produces a source classification as part of building the so-called \pastro quantity \cite{Abbott_2016_pastro}.
The classification aims to assess the origin of the signal by estimating the probability that it originated from a binary neutron star (BNS), neutron star-black hole (NSBH), binary black hole (BBH), or a non-astrophysical (terrestrial) source.
Source localization is done by running the Bayestar algorithm \cite{Singer_2016}, which uses as input the outcome of the matched filtering.

Since matched-filtering pipelines rely on a discretization of the parameter space, the results of the search may not yield the true SNR and identify the parameters best matching the signal. 
In the case of MBTA, the O4 template banks were designed to recover at least $96.5\%$ of the GW \snr{} across the searched space.
In certain regions, this minimal SNR recovery (commonly referred to as \textit{minimal match}) goes up to $98\%$ \cite{MBTAO4, MBTA_bank_note}.
The discreteness of the bank can lead to inaccuracies in the estimation of source parameters necessary to generate the source properties described above.
To mitigate this, PyCBC \cite{PyCBC1, PyCBC2} first deployed an online "\snropt{}" technique during the O3 run~\cite{PyCBC_SNRopt}, followed by GstLAL \cite{GstLAL1, GstLAL2, GstLAL3, GstLAL4} in O4a.
In this paper, we present an \snropt{} method based on the MBTA infrastructure, which was deployed during O4b. 
It performs a denser scan of the parameter space for sufficiently significant detections.
The new highest-\snr{} candidate is passed to the usual downstream analysis chain, including sky localization with Bayestar.
In parallel, we introduce a method that uses the outcome of the scan to derive posterior probability distributions for a subset of key intrinsic source parameters.
\\

In Section \ref{sec:02}, we describe the MBTA \snropt{} process. 
Section \ref{sec:03} explains how the results are used for fast estimation of the source's intrinsic parameters.
Section \ref{sec:04} presents the tests we conducted to evaluate the performance, along with the results obtained from applying the procedure to the recently released O4a LVK candidates \cite{GWTC4_results}.
We detail in Section \ref{sec:05} the \snropt{} configuration used by the MBTA pipeline during the second part of the O4c run.

\section{The MBTA \snropt} \label{sec:02}
In this section, we describe the various steps in the method used to determine the CBC parameters that best match the observed signal.

\subsection{Principle}

With the aim of improving the parameter recovery of sufficiently significant signals, we use \snr{} optimization to guide the search toward the most relevant region of the parameter space. 
The MBTA \snropt{} procedure first identifies this region based on the initial detection and covers it with a dense, localized template bank.
Matched filtering is then applied using a slightly different configuration compared to the original search.
In order to mitigate its dependence on the search results, this procedure is applied in a hierarchical manner to the data.

As the main challenge is to produce a better estimate of the source parameters in low latency, this must be done in a relatively short time, not exceeding a few minutes.
To keep the computational cost at a reasonable level, the procedure is applied only to triggers with a false alarm rate (\far) below one per two hours.

\subsection{Fast template placement}

The low-latency constraint prevents an initially too dense scan of the parameter space and requires methods that can be efficiently parallelized.
Information from the search can be exploited to restrict the mass and spin ranges where the source is most likely to lie.
Based on this, we have developed a method to rapidly place new templates, increasing the coverage of the parameter space surrounding the triggers initially recovered online.\\

Any CBC waveform $h$ is determined by a set of parameters $\boldsymbol{\lambda}$ which can be split into intrinsic $\boldsymbol{\lambda}_{\mathrm{int}}$ and extrinsic $\boldsymbol{\lambda}_{\mathrm{ext}}$ parameters.
MBTA templates are defined by the four dominant intrinsic parameters: detector-frame component masses $(m_1^d ,m_2^d )$ and the dimensionless spin components along the orbital angular momentum $(\chi_1^z,\chi_2^z)$.
The \textit{match} between two waveforms is defined as follows \cite{Owen_1996}:
\begin{equation}
    M(\boldsymbol{\lambda}_{\mathrm{int}}, \boldsymbol{\lambda}_{\mathrm{int}} +\Delta \boldsymbol{\lambda}_{\mathrm{int}}) = \max\limits_{\Delta\boldsymbol{\lambda}_{\mathrm{ext}}} \left< \bar{h}(\boldsymbol{\lambda}) \mid \bar{h}(\boldsymbol{\lambda} + \Delta \boldsymbol{\lambda}) \right> ,
    \label{02-eq:match}
\end{equation}
where waveforms are normalized such that $\bar{h} \equiv h / ||h||$, and the angular brackets denote the inner product, weighted by the noise one-sided power density spectrum (PSD) $S_n$ \cite{Cutler_1994}:
\begin{equation}
    <h_A \mid h_B> = 4 \Re{\int\limits_{0}^{+\infty} \frac{h_A(f)h_B^{*}(f)}{S_n(f)} df}.
    \label{02-eq:inner}
\end{equation}

The match quantifies the average fraction of \snr{} recovered by a given template relative to the true signal waveform.
To capture the variation of \snr{} across parameter space, we define a metric that quantifies how the match changes between two infinitesimally close points of coordinates $\boldsymbol{\lambda}$ and $\boldsymbol{\lambda} + \delta \boldsymbol{\lambda}$: 
\begin{equation}
    g_{\mu \nu} (\boldsymbol{\lambda}_{\mathrm{int}}) 
    \equiv - \left. \frac{1}{2} \frac{\partial^2 M(\boldsymbol{\lambda}_{\mathrm{int}}, \boldsymbol{\lambda}_{\mathrm{int}} + \delta \boldsymbol{\lambda}_{\mathrm{int}})}{\partial \delta \lambda_{\mathrm{int}}^{\mu} \partial \delta \lambda_{\mathrm{int}}^{\nu}} \right|_{\delta \boldsymbol{\lambda}_{\mathrm{int}} = 0},
    \label{02-eq:metric_def}
\end{equation}
which in practice is obtained by projecting onto the subspace orthogonal to extrinsic parameters the Fisher information matrix defined as follows \cite{Owen_1996, Owen_1999}:
\begin{equation}
    \Gamma_{ab}(\boldsymbol{\lambda}) = \left< \frac{\partial \bar{h}(\boldsymbol{\lambda})}{\partial \lambda^{a}} \left| \frac{\partial \bar{h}(\boldsymbol{\lambda})}{\partial \lambda^{b}} \right. \right>.
    \label{02-eq:fisher_matrix}
\end{equation}

With an appropriate choice of coordinates, the match between two nearby points can be well approximated using a second-order Taylor expansion, assuming the metric is locally constant:
\begin{equation}
    M(\boldsymbol{\lambda}_{\mathrm{int}}, \boldsymbol{\lambda}_{\mathrm{int}} + \Delta \boldsymbol{\lambda}_{\mathrm{int}}) \approx 1 - g_{\mu \nu}(\boldsymbol{\lambda}_{\mathrm{int}}) \Delta \lambda_{\mathrm{int}}^{\mu} \Delta \lambda_{\mathrm{int}}^{\nu},
    \label{02-eq:match_metric_approx}
\end{equation}
where repeated indices $\mu$ and $\nu$ are implicitly summed over.
For inspiral-dominated waveforms, such as BNS, a natural choice of parameters are the so-called post-Newtonian (PN) phasing coefficients \cite{PN_first, PN_last}.
Previous works \cite{Ajith_2011} have demonstrated that the leading-order terms (up to 2.5 PN order) can be fully characterized by the two component masses and a single reduced spin parameter:
\begin{equation}
    \chi_r = \left( 1 - \frac{76 \eta}{113} \right) \chi_s + \delta \chi_a,
    \label{02-eq:reduced_spin}
\end{equation}
where $\chi_{s/a} = \left(\chi_1^z \pm \chi_2^z\right)/2$ are the symmetric and asymmetric combinations of the component spin projections onto the orbital axis. 
The parameters $\eta = m_1 m_2 / \mtot^2$ and $\delta = (m_1 - m_2)/\mtot$ denote the frame-independent symmetric and anti-symmetric mass ratios, respectively, with $\mtot = m_1 + m_2$ the binary total mass.
Since higher-order terms are not expected to contribute significantly, we model our waveform templates using only these three parameters.
This choice is also computationally advantageous for online searches, as exploring a three-dimensional space is considerably less demanding than navigating the full intrinsic space of CBCs.
A convenient expression of these three parameters is given by the so-called dimensionless chirp times $\boldsymbol{\lambda}_{\mathrm{int}} = \boldsymbol{\theta} \equiv \left\{ \theta_0 , \theta_3 , \theta_{3s} \right\}$, in which the metric is expected to vary slowly across the parameter space  \cite{Ajith_2014}:
\begin{equation}
    \begin{split}
        & \theta_0 = \frac{5}{2^{1/3}} \left( \frac{G \Msun c^{-3}}{16 \pi f_0  \mchirp^d} \right)^{5/3}, \\
        & \theta_3 = \left( \frac{16 \pi^5 \theta_0^2}{25 \eta^3 } \right)^{1/5}, \\
        & \theta_{3s} = \frac{113 \chi_r \theta_3}{48 \pi},
    \end{split}
    \label{02-eq:chirp_times}
\end{equation}
where $\mchirp^d = \eta^{3/5} \mtot^d$ is the detector-frame chirp mass, $G$ is the gravitational constant, $\Msun$ the mass of the sun, $c$ the speed of light and $f_0$ an arbitrary frequency.
Following previous works \cite{Roy_2017, Roy_2019}, we extend the description of the metrics in this parameter space to CBCs up to higher masses.
\\

Rearranging Equation \ref{02-eq:match_metric_approx} as a function of the \textit{mismatch} ($\equiv 1 - M$), we can recognize the equation for an ellipsoid.
Therefore, in this framework, a surface of constant mismatch around a given template can be treated as a sphere of radius $R = \sqrt{1 - M}$, deformed by a linear transformation fully described by the matrix corresponding to the metric at that point.
Noticing that this sphere is universal and does not depend on the template under consideration, placing new templates around the ones that triggered the \snropt{} boils down to the following steps: 
\begin{itemize}
    \item pre-generating a \textit{universal} template bank in an abstract space where mismatches between points correspond to simple Euclidean distances, 
    \item projecting this bank onto the dimensionless chirp-time space using the metric of the parameter space at the location of the template that maximizes the posterior probability density function (PDF) (Equation \ref{03-eq:bayes}).
    \item and finally mapping the resulting dimensionless chirp times to the physical parameters \tripledash{} such as individual masses and spins \tripledash{} used to generate the waveforms for the matched filtering.
\end{itemize}

For the first step, we use a simple stochastic algorithm to generate independent samples distributed within a sphere.
This exact process is described in Appendix \ref{sec:A01_universalbank}.
A key point is the decision to accept or reject a new sample. 
This decision determines the bank’s minimal match.
We check that the candidate sample has no neighbor already accepted within a distance corresponding to the targeted minimal match.  
By adjusting this distance according to a given density function, we are able to fine-tune the coverage of the \snropt{} bank throughout the space.

The details of the second step are also given in Appendix \ref{sec:A01_universalbank}.
We follow the procedure outlined in \cite{Roy_2019} to compute the metric, using numerical estimates of the waveform derivatives in Equation \ref{02-eq:fisher_matrix}. 
Accounting for both the measured \snr{}s and astrophysical priors (described in Section \ref{sec:03}) in the choice of the bank center helps favor symmetrical binaries, which tend to be under-represented in the search bank.

The final step is to convert the dimensionless chirp times (Equation \ref{02-eq:chirp_times}) into individual detector-frame component masses ($m_1^d , m_2^d $) and spins ($\chi_1^z, \chi_2^z$). 
For the sake of simplicity, we distribute the reduced spin (Equation \ref{02-eq:reduced_spin}) equally between the two components:
\begin{equation}
    \begin{split}
        & \mtot^d = \frac{5}{32 \pi^2 f_0} \frac{\theta_3}{\theta_0} \frac{G \Msun}{c^3},  \\
        & \eta = \left( \frac{16 \pi^5}{25} \right)^{1/3} \frac{\theta_0^{2/3}}{\theta_3^{5/3}}, \\
        & \chi_r = \frac{48 \pi \theta_{3s}}{113 \theta_3}, \\
        & m_1^d = \frac{1}{2} \mtot^d \left( 1 + \sqrt{1 - 4 \eta} \right), \\
        & m_2^d = \frac{1}{2} \mtot^d \left( 1 - \sqrt{1 - 4 \eta} \right), \\
        & \chi_{1,2}^z = \frac{\chi_r}{1 - 76 \eta / 113}.
    \end{split}
    \label{02-eq:chirp_times_to_phys}
\end{equation}
This convention guarantees that the two aligned spins are equal to the so-called effective spin, defined as $\spineff = \left( m_1 \chi_1^z + m_2 \chi_2^z \right) / \mtot$.
Templates that do not correspond to physically valid masses or spins are simply discarded (see Appendix \ref{sec:A01_universalbank}).
\\

Figure~\ref{02-fig:bank_conversion_example} illustrates the template placement: the top row shows the universal samples, the middle row their projection into the dimensionless chirp time space using the metric estimated at the parameters marked by the purple star ($m_1^d = 2.9~\Msun$, $m_2^d = 1.9~\Msun$, $\spineff = -0.23$), and the bottom row the mass and spin parameters of the templates selected for filtering.

\begin{figure*}
    \centering
    \includegraphics[width=1\textwidth]{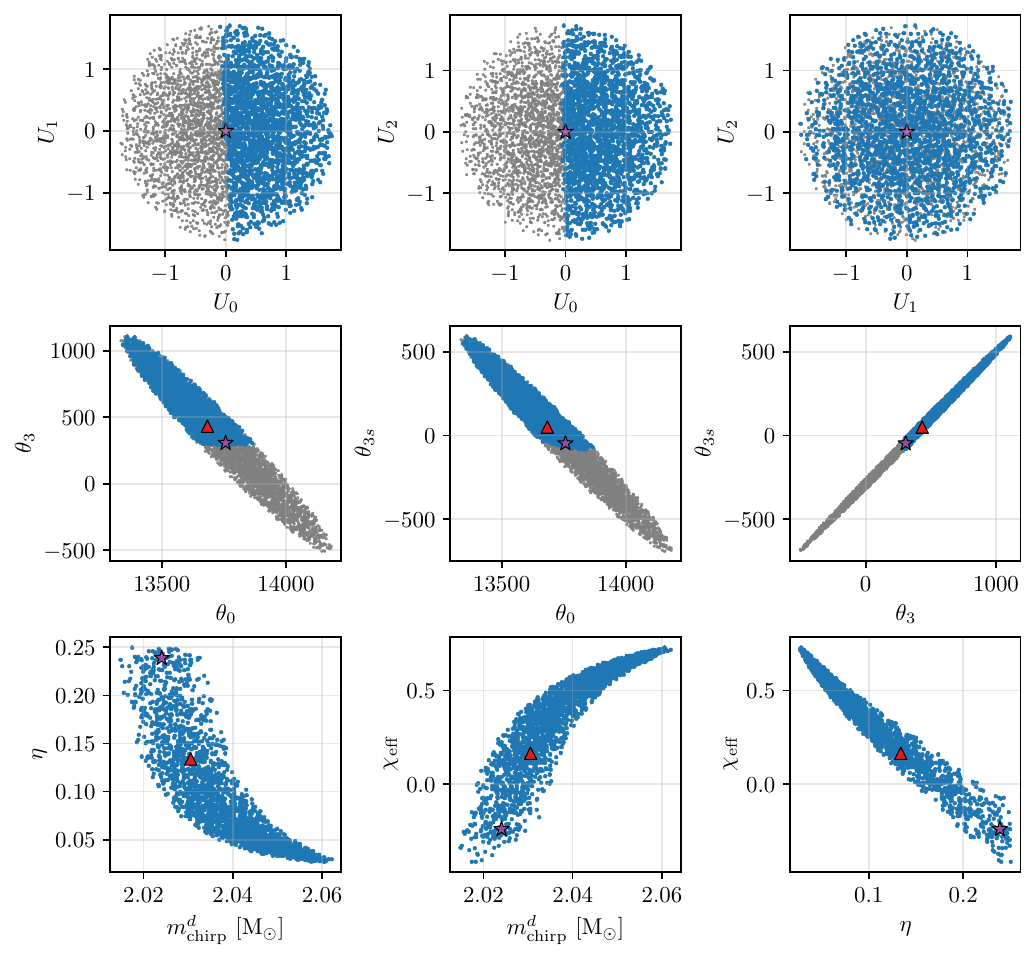}
    \caption{
    Illustration of the \snropt{} bank construction process for GW230529\_181500 \cite{Abac_2024}. 
    Each blue point represents a template. The purple stars indicate the bank center \tripledash{} i.e. the parameters of the most probable MBTA trigger for this event after accounting for astrophysical priors described in Section \ref{sec:03}. 
    The parameters recovered online by the highest \snr{} trigger from the MBTA pipeline are highlighted by red triangles.
    All values corresponding to non-physical parameters are grayed out in the top and middle rows, and excluded from the bottom row.
    The top row shows the 2D projections of the universal bank in the coordinate system defined by Equation~\ref{A01-eq:new_var}. 
    In this case, the templates are uniformly distributed in this space.
    The middle row displays the same samples transformed into dimensionless chirp time coordinates using Equation~\ref{A01-eq:U_to_theta} (with $\boldsymbol{S}=(1, 1, 2)$).
    The bottom row presents the final set of templates in a more commonly used parameter space: detector-frame chirp mass $\mchirp^d$ (in solar masses), symmetric mass ratio $\eta$ and effective spin $\spineff$.
    }
    \label{02-fig:bank_conversion_example}
\end{figure*}

\subsection{Filtering}

Once the bank has been determined, a matched filtering process \tripledash{} similar to that used in the MBTA search \cite{MBTAO4} \tripledash{} is applied. 
Templates with a total mass below $4 ~\Msun$ are generated using the \STTFOUR{} waveform model \cite{PhysRevD.90.124029}, while those with higher masses employ \SEOBNRFOUROPT{} to capture features beyond the inspiral \cite{Devine_2016}.
The filtering is performed in the frequency domain over the range from $24~\hertz$ ($20~\hertz$ for total mass above $30 ~\Msun$) to $2048~\hertz$.
It uses a single frequency band \tripledash{} unlike the multi-band approach adopted by most of the MBTA search, which requires a too long initialization time to be used for templates generated on the fly.
Since the template lengths are not known in advance, multiple data buffers ranging from $12$ to $200$ seconds are used for the filtering and PSDs estimation.
\\

As the MBTA \snropt{} is only triggered for candidate events and not for background, it does not attempt to re-assess their statistical significance.
Therefore, no additional noise rejection or veto is performed and all resulting triggers \tripledash{} along with their multi-detector combinations \tripledash{} are recorded, regardless of their \snr.
The highest \snr{} trigger is then uploaded to \gracedb, where it is used to update the MBTA SNR measurement and a new sky map is computed with the updated \snr{} time series.
The method used to update the source classification is discussed in Section \ref{sec:03}.
\\

To accelerate the computation, the filtering is parallelized across $160$ physical cores. 
However, this does not require additional computing resources, as the same cores used for online searches are employed, which typically operate at about $40\%$ load to maintain low latency.
The latency is only marginally impacted when running the \snropt{}, which occurs for about one minute every two hours.

\subsection{Hierarchical processing}

Because it focuses on a subspace that depends on the parameters initially recovered by the search, our method is inherently sensitive to the accuracy of those parameters.
However, online searches \tripledash{} optimized for SNR recovery \tripledash{} often misestimate source parameters, due to coarse template granularity (especially in the high-mass region), limited bank extent, and lack of proper prior.
Therefore, a significant bias in the online-estimated parameters may result in a suboptimal \snropt{} bank that fails to cover the space containing the true source parameters.
We mitigate this by applying the MBTA \snropt{} in a hierarchical manner.

A first pass is carried out using an agnostic template distribution, sampled uniformly across the dimensionless chirp time space.
The purpose of this step is to probe regions of the parameter space that may have been neglected by the search, either due to boundary limitations or non-uniformities in its template bank.
A minimal match of $98\%$ is employed.
Since Equation \ref{02-eq:match_metric_approx} tends to underestimate the match, the resulting \snropt{} bank is denser than expected, often yielding a gain in SNR at this stage. 
Simulations show that a few percent of sources may fall outside the bank, likely due to poorly measured spins and inability to properly take into account fast metric variations along certain dimensions.
To better probe these dimensions, we apply a scaling factor of $2$ along the third universal dimension when projecting the samples (see Equation \ref{A01-eq:U_to_theta}). 
This mainly affects the values of $\theta_{3s}$ \tripledash{} which governs the relation between mass ratios and spins \tripledash{} of the new templates, with little impact on chirp mass recovery. 
This scaling is compensated by generating a denser bank to maintain the desired minimal match.
This bank is designed to hold a maximum of $4000$ templates.

In parallel with sending out the highest SNR trigger recovered from the first pass, all triggers produced undergo the procedure outlined in Section \ref{sec:03}.
The result is an estimate of the most probable detector-frame source intrinsic parameters, which is used to initiate a second iteration of the \snropt{}.
The bank for this stage is centered on the newly identified parameters, with the metric recomputed accordingly.
This second pass aims to refine the source parameter estimation.
Given that the likelihood expressed in Equation \ref{03-eq:likelihood} scales exponentially with the SNR difference, we sample the new bank such that the minimal match decreases exponentially away from the bank center, as detailed in Appendix \ref{sec:A01_universalbank}.
This second bank contains up to $7000$ templates.
\\

For consistency, the template yielding the highest recovered SNR at each step is added to the bank for the next iteration.
The \snropt{} employs a slightly different method for estimating the PSD from the MBTA search, but then ensures that both \snropt{} iterations use the same PSD.

\section{Parameter estimation} \label{sec:03}
Given the limited observational resources, the primary challenge for EM counterpart searches is to determine which events are most relevant to follow up.
To support this decision-making process, low-latency CBC searches must provide characterizations of their candidates. 
This is typically achieved by reporting source classification and properties, expressed as numerical probabilities that the source meets specific criteria \cite{UserGuide}.
Misidentification can result in missing potentially EM-bright candidates or, conversely, initiating resource intensive follow-up observations for less promising events. 
Recently, chirp mass estimates have been added to the information released in LVK alerts \cite{Chaudhary_2024, UserGuide}. 
This parameter is particularly valuable for predicting the remnant's light curve \cite{Margalit_2019, Toivonen_2025}.

Because search pipelines rely on a discretized parameter space and make numerous simplifications, they often misestimate source parameters. 
More computationally intensive Bayesian inference methods can be used, such as the Bilby pipeline \cite{Ashton_2019}, to explore the full parameter-space dimensionality with a variety of waveform approximants and obtain precise and unbiased estimates.
While traditional full PE approaches can require hours to days, recent and ongoing developments aiming to deliver results within minutes \cite{Pankow_2015, rose2022, Morisaki_2023}.

In this section, we explore the possibility of exploiting the MBTA \snropt{} results to quickly infer key intrinsic source properties by focusing on the parameter space dimensions explored by the MBTA search.

\subsection{MBTA detector-frame PE}

The \snropt{} procedure presented here performs a dense scan of the parameter space.
Assuming the filtered template bank is sufficiently dense, we can use the measured \snr{} to extract source parameters. 

Under the hypothesis that a CBC signal is present in the data $d$, the posterior PDF for the parameters $\boldsymbol{\lambda}$ of a given template $T$ is given by Bayes theorem, as the product of the template likelihood and prior PDF:
\begin{equation}
    f(T(\boldsymbol{\lambda}) \mid d) \propto \mathcal{L}(d \mid T(\boldsymbol{\lambda})) \cdot f(T(\boldsymbol{\lambda})).
    \label{03-eq:bayes}
\end{equation}

Assuming statistically independent detector noise, the likelihood factorizes across detectors. 
Under the additional assumption of stationary Gaussian noise, and after maximizing over the extrinsic CBC parameters ($\boldsymbol{\lambda}_{\mathrm{ext}}$), the likelihood can be expressed as \cite{McWilliams_2010}:
\begin{equation} 
    \mathcal{L}(d \mid T({\boldsymbol{\lambda}_{\mathrm{int}}})) 
    \propto \exp \left( - \sum\limits_i \snr_{\mathrm{opt},i}^2 \left( 1 - \frac{\snr_{T,i}}{\snr_{\mathrm{opt},i}} \right) \right),
    \label{03-eq:likelihood} 
\end{equation} 
where $i$ denotes the detector index. 
The term $\snr_{T,i}$ is the matched-filter \snr{} of template $T$ in detector $i$, while $\snr_{\mathrm{opt},i}$ denotes the (unknown) true signal amplitude. 
We approximate $\snr_{\mathrm{opt},i}$ by the single-detector \snr{} associated with the template yielding the maximum network \snr, in order to ensure robustness against local noise fluctuations.

The second term of Equation \ref{03-eq:bayes} is the prior PDF of the source parameters.
To ensure consistency with the parameters reported by the collaboration in \cite{GWTC4_results}, we adopt the same agnostic priors: uniform distributions in detector-frame individual masses and spin magnitudes, with isotropic spin orientations.
This prior is expressed in the dimensionless chirp time space following the procedure described in Appendix \ref{sec:A02_priorcalculation}.
\\

\begin{figure}[h]
    \centering
    \includegraphics[width=0.9\linewidth]{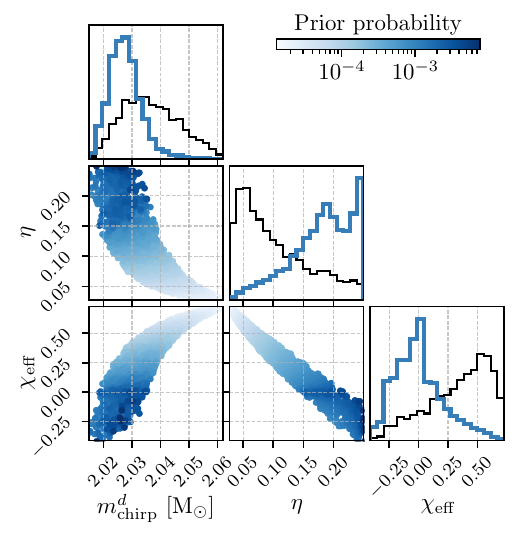}
    \caption{Prior probability distribution of the templates used in the first iteration of the \snropt{}, initiated by the MBTA trigger for GW230529\_181500 \cite{Abac_2024}, used alongside recovered SNRs to infer source parameters.
    The distributions are shown for the detector-frame chirp mass ($\mchirp^d$), the symmetric mass ratio ($\eta$), and the effective spin ($\spineff$).
    The plots display the marginal 1D (diagonal) and 2D (off-diagonal) distributions.
    Black lines indicate the distributions of the templates used for filtering.
    Blue lines and dots show the same distributions reweighted according to Equation \ref{A02-eq:prior_calculation}.
    }
    \label{03-fig:prior_example}
\end{figure}

To account for the fact that our samples are drawn from the \snropt{} template bank, and therefore follow a specific distribution in dimensionless chirp times, we apply the reweighting procedure described in \cite{Chatziioannou_2024}, resutlting in probability distributions obtained by dividing the sampled PDFs by the template density \tripledash{} either uniform for the first iteration or following Equation \ref{A01-eq:template_density_it2} for the second iteration.
Figure \ref{03-fig:prior_example} highlights the differences between the template distribution and the prior probability distribution, illustrating that the chosen prior favors symmetric masses and low effective spins, while the bank covers a broader parameter space.
Figure \ref{03-fig:post_GW230529_detector_frame} shows the detector-frame posterior probability distribution evaluated for the parameters of each template filtered during the second iteration of the MBTA \snropt{}, applied to the data associated with GW230529\_181500 \cite{Abac_2024}.

\subsection{MBTA source-frame PE}

The source properties can be inferred from mass and spin estimates given by Equation \ref{03-eq:bayes}.
However, because of the Universe's expansion, the masses measured in the detector-frame appear larger due to redshift $z$ \cite{1987GReGr..19.1163K}.
Since our PE method only probes the intrinsic parameters, it does not infer the source distance, nor the redshift.
We therefore rely on the source localization provided in low-latency by the Bayestar algorithm \cite{Singer_2016}, applied to the maximum likelihood template. 
Following the procedure in Appendix \ref{sec:A03_frameconversion}, we convert the detector-frame mass and spin joint posterior probability distribution into source-frame using the marginal luminosity distance probability distribution provided by Bayestar and assuming a standard cosmological model, described in \cite{Planck2015}.
\\

\begin{figure}
    \centering
    \subfloat[Detector-frame. Samples are weighted by their posterior probability (shown in blue).]{\includegraphics[width=0.9\linewidth]{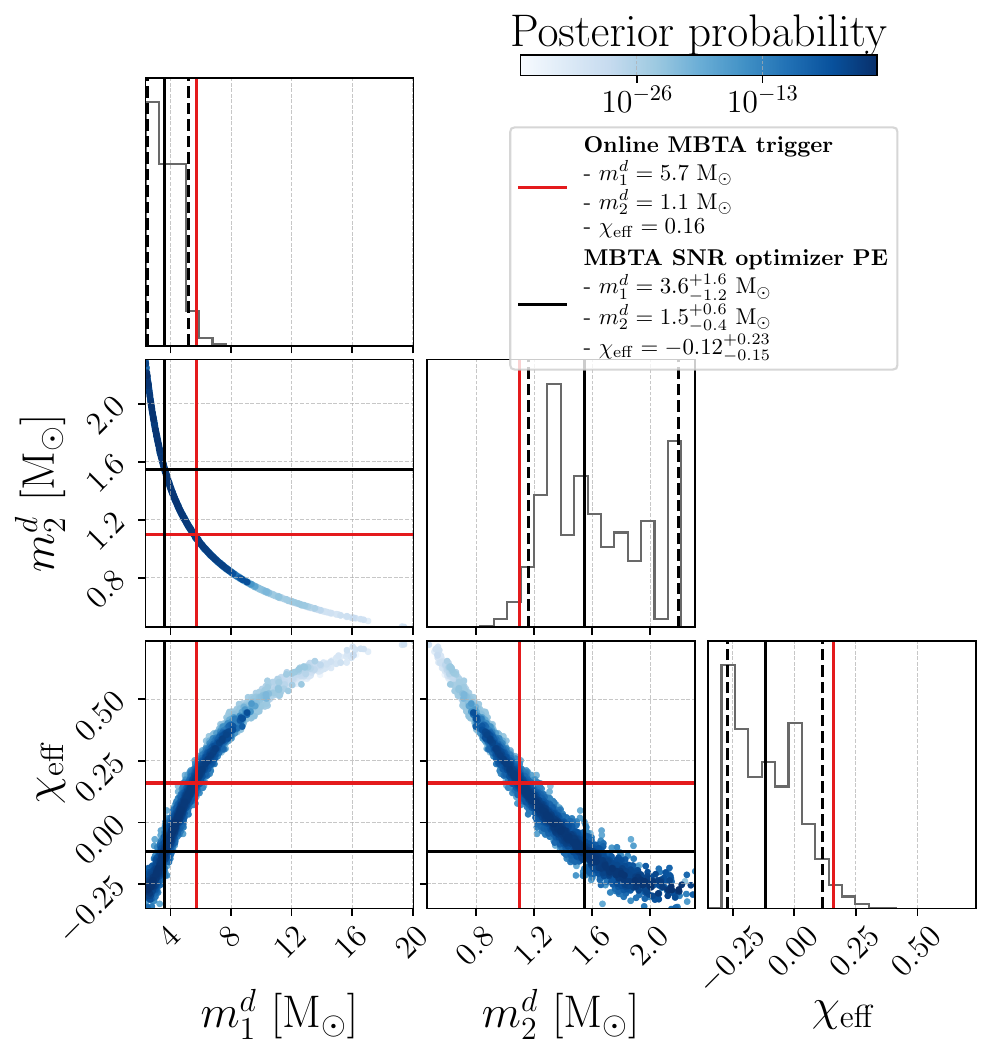}\label{03-fig:post_GW230529_detector_frame}} \\
    \subfloat[Source-frame. Samples are obtained by rejection sampling of the calculated posterior probability distribution.]{\includegraphics[width=0.9\linewidth]{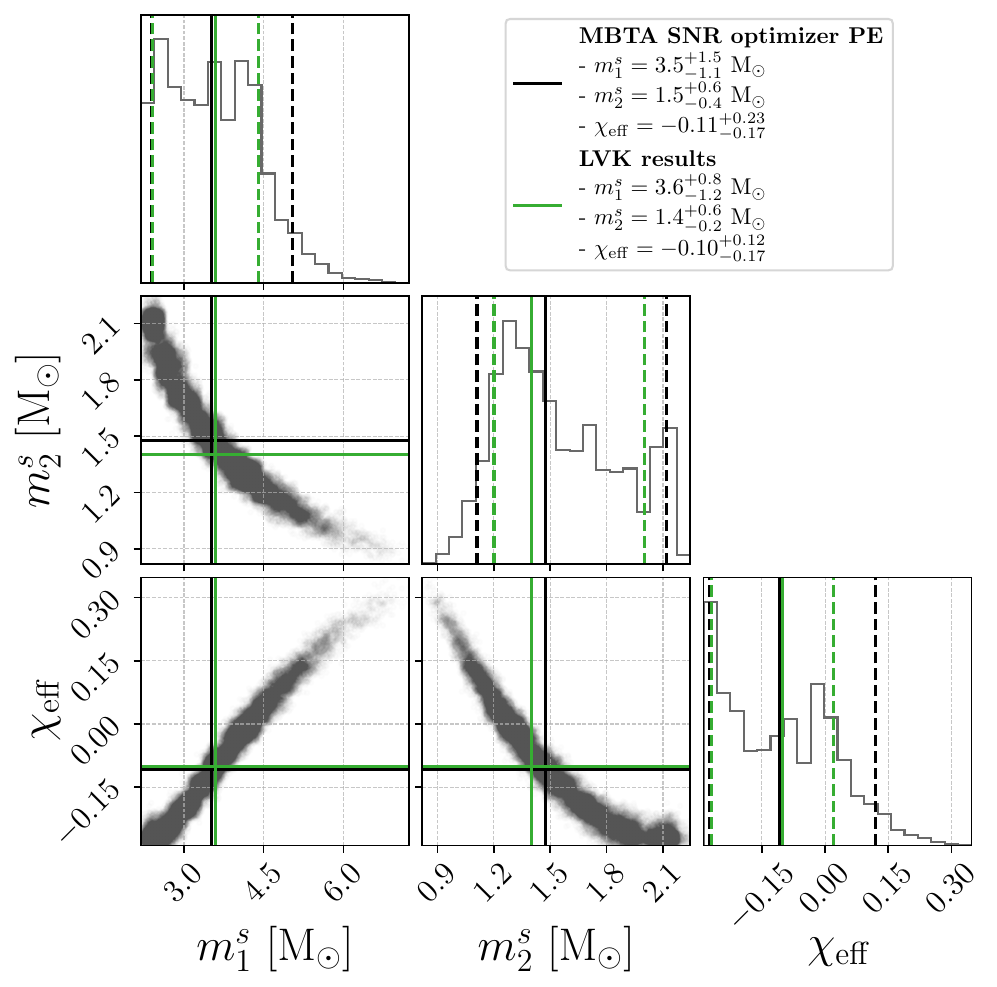}\label{03-fig:post_GW230529_source_frame}}
    \caption{Visualization of the parameter inference obtained after the second iteration of the MBTA \snropt{} on data associated with GW230529\_181500.
    The plot shows the 1D (diagonal) and 2D (off-diagonal) marginal posterior distributions
    for the component masses ($m_1, m_2$) and effective spin (\spineff).
    Red lines indicate the detector-frame parameters of the online most significant MBTA trigger.
    Black lines give the median (solid) and $90\%$ symmetric credible intervals (dashed) of the source parameters obtained by the MBTA \snropt{}.
    Green lines indicate the values inferred by the LVK collaboration in \cite{Abac_2024}.}
    \label{03-fig:post_GW230529}
\end{figure}

Figure \ref{03-fig:post_GW230529_source_frame} shows the results of this conversion for GW230529\_181500 and compares them with the values reported in \cite{Abac_2024}.
They give close results.

\subsection{Source classification and properties}

The resulting source-frame posterior probability distribution can be used to update the source classification and properties publicly provided to aid for inform EM surveys of the candidates.

Source classification is expressed as a set of probabilities representing the possible origin of the signal. 
Four mutually exclusive categories are considered: \pterr for a non-astrophysical (terrestrial) origin, \pbns for both component masses between $1~\Msun$ and $3~\Msun$, \pnsbh for a binary with one mass in this range and the other above $3~\Msun$, and \pbbh for both components above $3~\Msun$ \cite{UserGuide}.
Since the MBTA \snropt{} relies on a significant detection and does not re-estimate candidate significance, the value of \pterr is taken from the online search \cite{Andres_2022, MBTAO4}.
By integrating the source-frame joint individual mass posterior probability distribution, marginalized over spin, the remaining probability is redistributed among the three astrophysical categories.
Since standard LVK searches do not include a \pastro sub-solar mass (SSM) category, the portion of the posterior probability distribution corresponding to SSM components is excluded from this source classification process.

The collaboration also provides a set of probabilities inferring the possible properties of the source \cite{UserGuide}.
While the method for estimating these properties is shared across all LVK search pipelines \cite{Chaudhary_2024}, we show that the MBTA \snropt{} PE process can be used to produce our own estimate.
The \hasssm and \hasmg probabilities can be computed by directly integrating the source-frame mass posterior probability distribution over the regions where at least one component has a mass below $1~\Msun$, and between $3~\Msun$ and $5~\Msun$, respectively.
The \hasns and \hasrem definitions are dependent on neutron star models.
Therefore, the limits for neutron star and remnant masses are marginalized over several neutron star equations of state \cite{legred_2022_6502467} to weight the source-frame posterior probability distribution before integration.
In this process, neutron star spins in NSBH systems are fixed to zero, while black hole spins are allowed to vary according to the measured posterior probability distribution.

\section{Results} \label{sec:04}
To validate the MBTA \snropt{} and PE procedures, we constructed and analyzed several dedicated data sets.
We built two of them from the simulated CBC injections described in \cite{O4a_common_inj}.
These "O4a LVK common injections", used by all LVK search pipelines contributing to GWTC-4.0 \cite{GWTC4_results}, provide sufficient statistics for quantitative performance studies.
We retain only injections associated with a MBTA trigger with a FAR lower than one per year.

We define the \popone injection set as a subset of these injections. 
To mitigate the limited statistics for low-mass systems, the BNS and NSBH populations are oversampled by a factor of ten relative to their nominal abundance. 
The resulting data set comprises $160$ BNS, $133$ NSBH, and $707$ BBH injections. 
It is used to assess the improvements in SNR, sky localization, source classification and properties achieved with the MBTA \snropt{}.

In addition, we construct a second injection set, referred to as \poptwo, also derived from the O4a LVK common injections.
We resample the recovered injections so that the resulting population matches the priors used in our PE procedure.
This results in a set of $304$ prior-consistent injections suitable for evaluating the performance of our PE procedure.

Finally, to assess the agreement of our PE results with those obtained by the LVK collaboration, we also applied our PE method to the $71$ candidate events reported by the LVK collaboration in Table 3 of \cite{GWTC4_results}, for which MBTA identified a trigger online, regardless of its significance.
This set will be referred to as the \OFouraevents.

\subsection{SNR}

\begin{figure}[!h]
    \centering
    \subfloat[First \snropt{} iteration]{\includegraphics[width=\columnwidth]{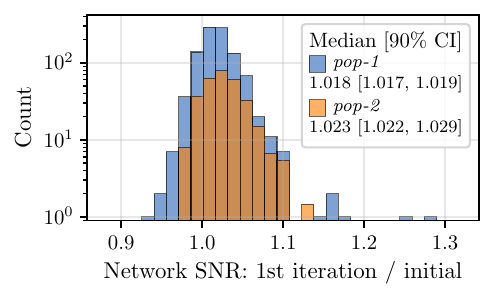}\label{04-fig:Inj_SNRGain_it1}} \\
    \subfloat[Second \snropt{} iteration]{\includegraphics[width=\columnwidth]{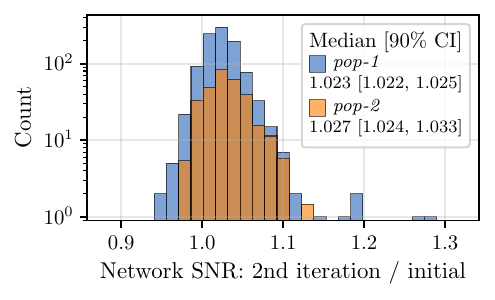}\label{04-fig:Inj_SNRGain_it2}}
    \caption{Distribution of the network \snr{} ratio between the \snropt{} output and the initial trigger from the search, evaluated on the two subsets of O4a LVK common injections.
    The $90\%$ symmetric credible intervals (CI) of the median are estimated using a bootstrap method.}
    \label{04-fig:Inj_SNRGain}
\end{figure}

A primary ingredient of the procedure presented in this paper is to increase the SNR measurement made by MBTA.
Figure \ref{04-fig:Inj_SNRGain} shows the gain of network \snr{} we measured on the \popone and \poptwo injection sets.
We observe that most of the SNR improvement comes from the first iteration, indicating that a more uniform template coverage \tripledash{} free from bank boundaries \tripledash{} enhances SNR, even though the effective coverage remains comparable to that of the search bank. 
After reanalyzing the data with a second iteration, the median SNR gain is approximately $2.3\%$ for the \popone set and $2.7\%$ for the \poptwo set.
This is consistent with the fact that the \popone set, which contains a high proportion of BNS and symmetric BBH systems, has a substantial fraction of sources within the $98\%$ minimal-match region of the search template bank, whereas the remaining sources \tripledash{} more prevalent in the \poptwo set \tripledash{} lie in the $96.5\%$ coverage region \cite{Andres_2022, MBTA_bank_note}.

In around $10\%$ of the cases, the \snropt{} procedure results in a slight reduction in the candidate’s \snr.
Since each iteration retains the loudest template from the previous one, this must be attributed to the way the data are filtered.
While MBTA typically employs a multi-band approach, the \snropt{} processes the data over the entire frequency band.
As this can involve templates with highly variable durations, we have to recompute the data FFTs and noise PSDs, resulting in random \snr{} fluctuations. 

\subsection{Sky maps}

\begin{figure}[!h]
    \centering
    \includegraphics[width=\columnwidth]{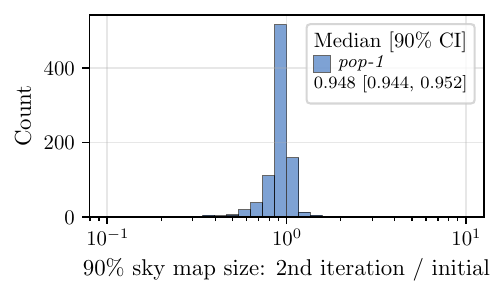}
    \caption{Distribution of the $90\%$ credible region ratio between the second \snropt{} iteration highest-SNR trigger and the initial trigger, evaluated on a subset of O4a LVK common injections (\popone).
    The $90\%$ symmetric credible intervals (CI) of the median are estimated using a bootstrap method.}
    \label{04-fig:SkyMapSize}
\end{figure}

\begin{figure}[!h]
    \centering
    \includegraphics[width=\columnwidth]{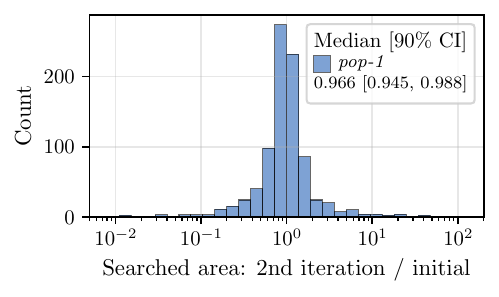}
    \caption{Distribution of the searched area ratio between the second \snropt{} iteration highest SNR trigger and the initial trigger, evaluated on a subset of O4a LVK common injections (\popone).
    The $90\%$ symmetric credible intervals (CI) of the median are estimated using a bootstrap method.}
    \label{04-fig:SkyMapArea}
\end{figure}

An expected outcome of the \snropt{} is to improve the candidate sky localizations, as the refined intrinsic parameters lead to improved extrinsic parameters.
We evaluate this by applying the Bayestar algorithm \cite{Singer_2016} to our \popone injection set.
We exclude $26$ single-detector triggers since their sky maps depend only on the detector antenna pattern and are independent of the SNR or any template intrinsic parameters.
Since the LVK collaboration only releases online information for the highest-\snr{} trigger, we focus on the $894$ coincident events for which a gain in \snr{} is observed.
Figure \ref{04-fig:SkyMapSize} compares the $90\%$ credible regions of the sky maps from the initial candidates with those generated with the output of the full \snropt{} procedure, selecting the highest-\snr{} trigger of the second iteration.
The observed $\sim 5\%$ reduction in sky map sizes for two-detector coincident events is consistent with a gain of about $2.5\%$ in \snr{} from each detector.
To quantify the accuracy of the localization, we also consider the searched area, defined as the smallest region of the sky, ranked by probability, that contains the true source location.
Figure \ref{04-fig:SkyMapArea} compares the searched areas corresponding to the initial candidates with those from the second iteration.
Although the map sizes typically change by a few tens of percent, the searched areas often differ by orders of magnitude.
We observe a significant reduction in the median searched area.
The median $50\%$ ($90\%$) sky map area decreases from $374~\mathrm{deg^2}$ ($1401~\mathrm{deg^2}$) for the initial trigger to $352~\mathrm{deg^2}$ ($1330~\mathrm{deg^2}$) for the highest-\snr{} trigger obtained in the second iteration of the \snropt{}. 
Similarly, the median searched area is reduced from $307~\mathrm{deg^2}$ to $294~\mathrm{deg^2}$.

\subsection{Parameter recovery}

\begin{figure}[!h]
    \centering
    \subfloat[\popone]{\includegraphics[width=0.9\linewidth]{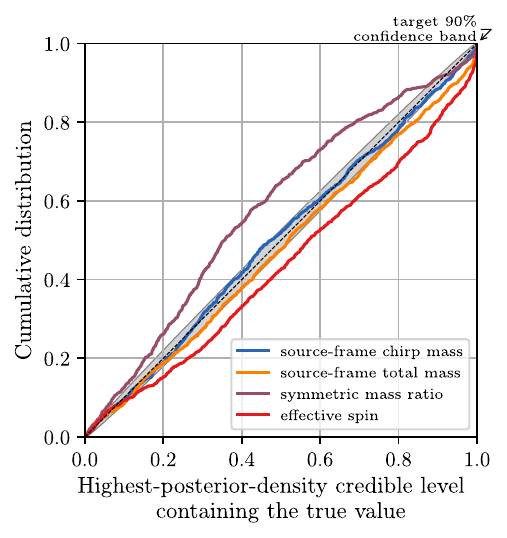}\label{04-fig:pp_plots_pop1}} \\
    \subfloat[\poptwo]{\includegraphics[width=0.9\linewidth]{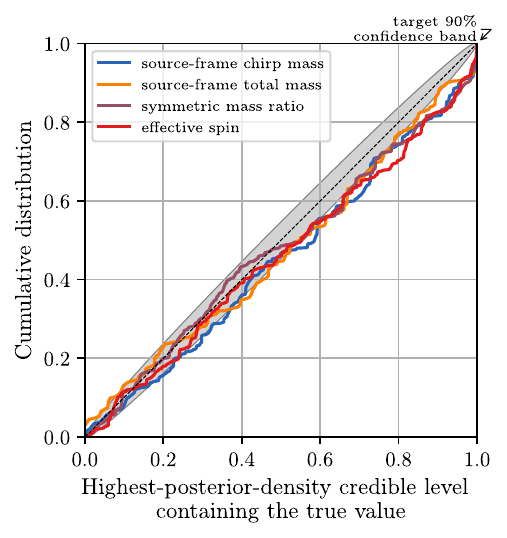}\label{04-fig:pp_plots_pop2}}
    \caption{
    P–P plots for key parameters recovered by the MBTA PE pipeline for the two subsets of O4a LVK common injections.
    The black dashed line represents the diagonal, indicating unbiased posterior distributions with optimal width, while the grey area shows the $90\%$ confidence band derived from the beta distribution.
    }
    \label{04-fig:pp_plots}
\end{figure}

Recently, the LVK collaboration began releasing low-latency estimates of the source chirp mass in the form of coarse-grained histograms. 
These estimates rely on detector-frame point estimates provided by the search pipelines. 
We aim to demonstrate that our PE procedure is effective at measuring source parameters and producing such histograms.
Figures \ref{04-fig:pp_plots_pop1} and \ref{04-fig:pp_plots_pop2} show probability–probability (P–P) plots for parameters of interest, constructed from the results of the second \snropt{} iteration applied to the \popone and \poptwo injection sets, respectively. 
These plots indicate that the posterior distributions of the source-frame chirp and total masses are consistent with the true values and are robust to the choice of prior.
In contrast, Figure \ref{04-fig:pp_plots_pop1} shows deviations for the symmetric mass ratio and effective spin in \popone, whereas no such deviations are observed in Figure \ref{04-fig:pp_plots_pop2}, where \poptwo follows the assumed prior. 
This suggests that the discrepancies in \popone likely result from the subtle impact of these parameters on the waveforms, which makes their estimation more sensitive to the choice of prior.
In addition, small deviations from the diagonal at high quantiles are observed for all parameters in both P–P plots. 
In rare cases, this restriction can exclude the true signal parameters.
These deviations may arise from the restriction of the explored parameter space based on the initial detection, which in rare cases can exclude the true signal parameters, and from limitations due to template placement based on a constant metric at high mass, which does not fully capture asymmetries in the match variation when merger and ringdown contributions become significant.
We find that this effect remains negligible for total masses below $100 ~\Msun$, but above, it introduces a bias toward lower masses of order $5\%$ for $m_{tot}^d \sim 200 ~\Msun$.
This has limited impact on the inferred source classification and properties, as such high-mass systems are easily identified as BBH and accurate low-latency PE is usually less critical for them.

\subsection{Comparison with PE results reported in GWTC-4.0}

For further validation, we applied our PE method to the \OFouraevents and compared the posterior probability distributions obtained with our method to those provided by the LVK collaboration.
We perform this comparison by evaluating the overlap $\mathcal{O}$ between the two binned distributions \tripledash{} denoted $p_{\mathrm{MBTA}}$ and $p_{\mathrm{LVK}}$ \tripledash{} given by \cite{Bhattacharyya}:
\begin{equation}
\mathcal{O}(p_{\mathrm{MBTA}}, p_{\mathrm{LVK}}) = \sum\limits_i \sqrt{\bar{p}_{\mathrm{MBTA}}(i)  \bar{p}_{\mathrm{LVK}}(i)},
\label{04-eq:overlaps}
\end{equation}
where $\bar{p} = p / \sum\limits_i p(i)$ is the normalized probability, and the sum is over all bins $i$\footnote{We use $20$ bins, having checked that the results are stable using $15$ to $30$ bins.}.

\begin{figure}[!h]
    \centering
    \subfloat[Source-frame individual masses (2D overlaps)]{\includegraphics[width=0.9\linewidth]{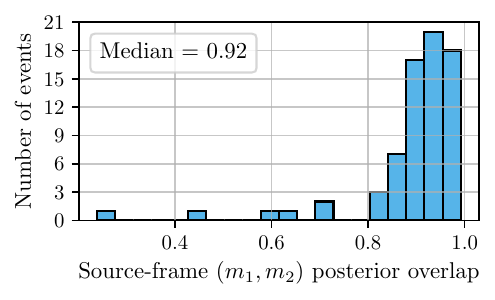}\label{04-fig:O4aevents_overlaps_m1_m2}} \\
    \subfloat[Effective spin (1D overlaps)]{\includegraphics[width=0.9\linewidth]{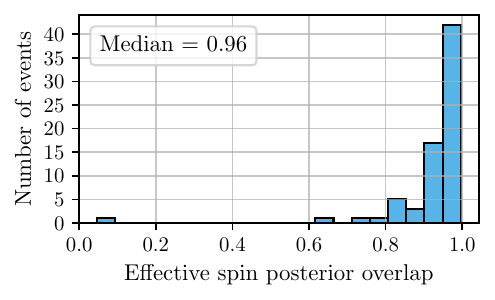}\label{04-fig:O4aevents_overlaps_seff}}
    \caption{
    Distribution of overlaps between MBTA and LVK posterior distributions, (defined in Equation~\ref{04-eq:overlaps}) for GW candidates reported in Table 3 of \cite{GWTC4_results}, for which MBTA identified a trigger, regardless of its significance (\OFouraevents).
    }
    \label{04-fig:O4aevents_overlaps}
\end{figure}

Figure \ref{04-fig:O4aevents_overlaps_m1_m2} shows the overlap distribution of the $71$ events in the dataset for the 2D source-frame component masses posterior distributions.
Most events are in qualitative agreement with the LVK results\footnote{The LVK results are obtained using full Bayesian PE, also using different waveform models, data calibration and noise mitigation.
Therefore, perfect agreement is not expected.}, with only six cases (all BBH candidates) exhibiting an overlap below $80\%$:
\begin{itemize}
    \item GW230709\_122727 ($\mathcal{O} = 68\%$), GW231001\_140220 ($\mathcal{O} = 29\%$) and GW240107\_013215 ($\mathcal{O} = 49\%$) are events for which the MBTA PE procedure inferred consistent source-frame total masses, but more extreme mass ratios compared to the LVK collaboration.
    This discrepancy is likely due to the presence of non-Gaussian artifacts in the data from one of the detectors.
    In the LVK analysis, this effect was mitigated by adjusting the starting frequency of the analysis \cite{GWTC4_results}, whereas our procedure relies on online data without in-depth noise mitigation.
    Running the MBTA \snropt{} with the same starting frequency used in the LVK analysis increases the overlaps to $99\%$, $84\%$ and $77\%$, respectively.
    Running extra iterations with different starting frequencies to check the stability of the results may be implemented in the future.
    \item GW230820\_212515 ($\mathcal{O} = 58\%$) is a marginally significant candidate ($\far \gtrsim 1/4~\peryr$) \cite{GWTC4_results}.
    Although the LVK spin posterior favors a positive effective spin, the MBTA PE reports a zero-spin result (which remains consistent with the LVK confidence interval).
    This leads to an asymmetry of the component masses in the MBTA inference.
    \item GW230922\_040658 ($\mathcal{O} = 69\%$) is a high-mass BBH candidates ($m_{\mathrm{tot}}^s \sim 125~\Msun$) observed at a redshift of around $1$ \cite{GWTC4_results}, one of the largest measured at that time for a GW event.
    The MBTA PE infers detector-frame masses consistent with the LVK results, but the source luminosity distance is underestimated, leading to larger source-frame masses compared to the LVK.
    This behavior is not expected for lower-mass CBC events, where the redshift is better constrained and has a smaller impact on the source-frame PE.
    \item GW231123\_135430 ($\mathcal{O} = 65\%$) is the most massive BBH confidently observed with GW to date ($m_{\mathrm{tot}}^s \sim 240~\Msun$) \cite{GW231123_135430, GWTC4_results}.
    It exhibits significant systematic differences in the inferred parameters when analyzed with different waveform models \cite{GW231123_135430, GWTC4_results}.
    It is therefore not surprising that our PE method, which employs a less exhaustive model (\SEOBNRFOUROPT \cite{Devine_2016}) than the standard LVK analysis, produces different results.
    Again, this behavior is not expected for low-mass events.
\end{itemize}

Figure \ref{04-fig:O4aevents_overlaps_seff} shows the overlap distribution of the 1D effective spin posterior distributions.
Overall, we find high overlaps with the LVK results, with only three outliers \tripledash{} GW231001\_140220 ($\mathcal{O} = 62\%$), GW240107\_013215 ($\mathcal{O} = 72\%$) and GW231123\_135430 ($\mathcal{O} = 5\%$) \tripledash{} which are a subset of the mass-estimate outliers discussed above.

\begin{figure*}
    \centering
    \includegraphics[width=1\textwidth]{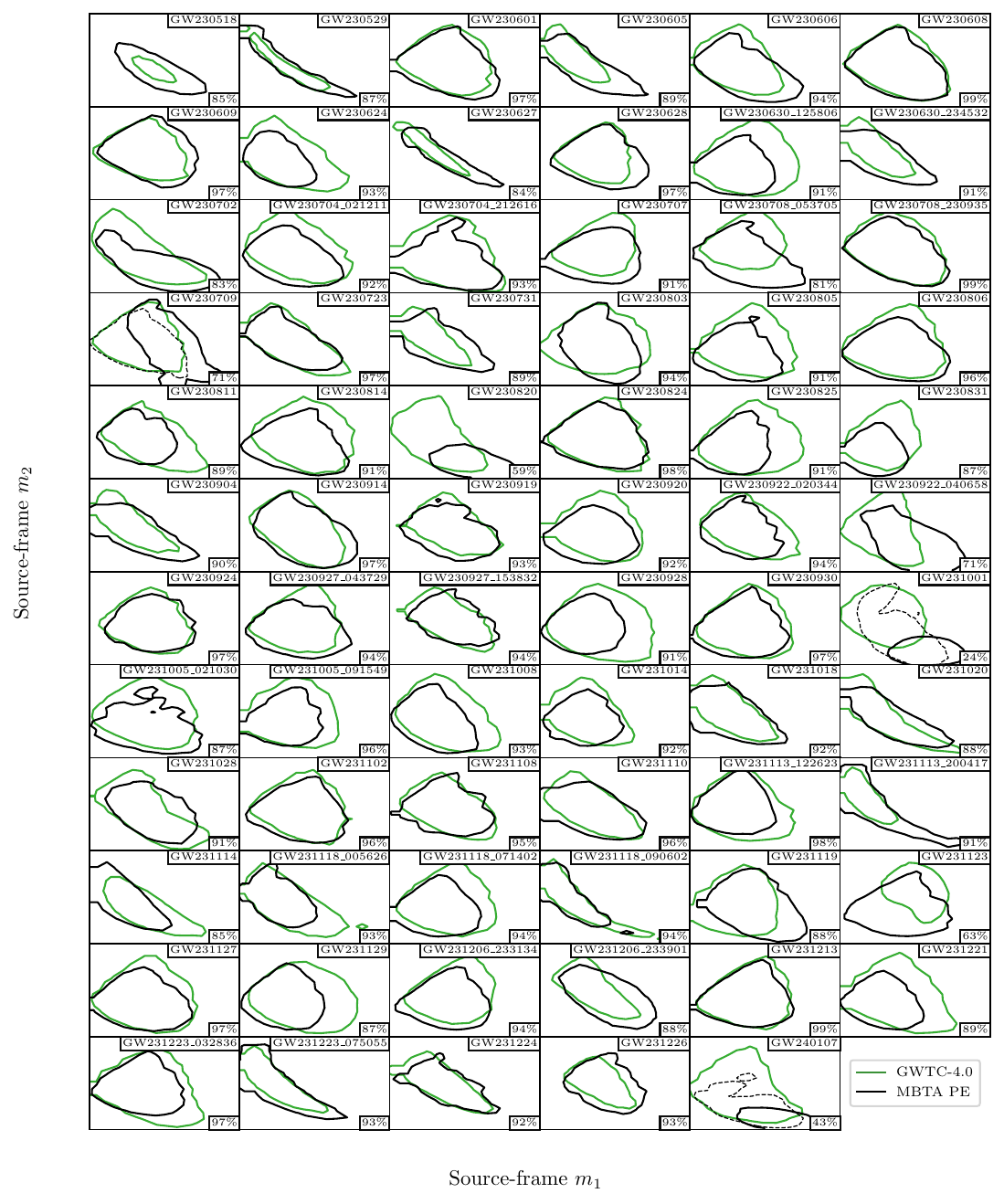}
    \caption{
        Comparison of $90\%$ symmetric credible intervals for source-frame component masses between LVK (green) and MBTA \snropt{} PE results (black).
        Dotted black lines show the results obtained using the same starting frequency as the LVK analysis for GW230709\_122727, GW231001\_140220, and GW240107\_013215 to mitigate non-Gaussian noise.
        Each subplot corresponds to a GW candidate event reported in Table 3 of \cite{GWTC4_results}, for which MBTA has identified a trigger, regardless of its significance (\OFouraevents).
        LVK posterior contours are obtained by combining equal numbers of samples \cite{GWTC4_zenodo} from each filtered waveform model listed in \cite{GWTC4_results}.
        For readability, event name suffixes have been omitted when unambiguous.
        The numbers in the bottom-right corners indicate the overlap between the two posterior distributions, as defined in Equation \ref{04-eq:overlaps}.
    }
    \label{04-fig:O4aevents_mosaic}
\end{figure*}

Figure \ref{04-fig:O4aevents_mosaic} shows a visual comparison of the $90\%$ symmetric credible intervals for each event. 
The two posterior sets indicate that the MBTA PE provides reliable source parameter estimates for most events.

\subsection{Source classification and properties}

\begin{figure}[!h]
    \centering
    \subfloat[Initial classification]{\includegraphics[width=0.9\linewidth]{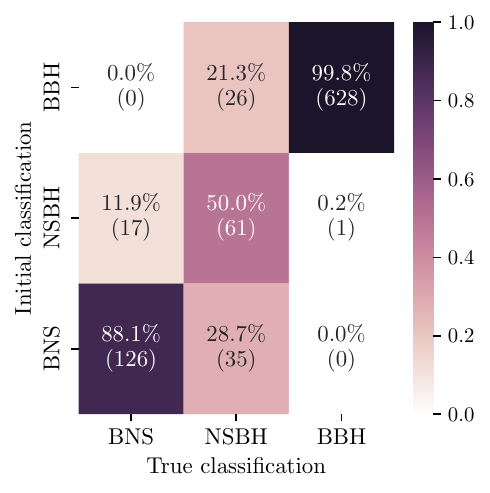}\label{04-fig:ClassificationInitial}} \\
    \subfloat[2nd iteration classification]{\includegraphics[width=0.9\linewidth]{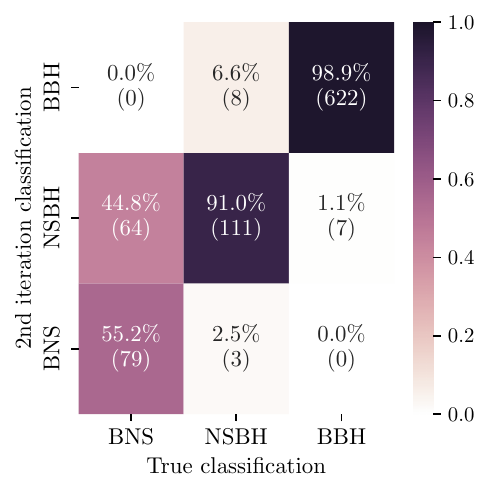}\label{04-fig:ClassificationIt2}}
    \caption{Confusion matrices of source classification for a subset of O4a LVK common injections (\popone). 
    Each injection is assigned to its most probable class.}
    \label{04-fig:Classification}
\end{figure}

By applying our PE procedure to the \snropt{} output, we are able to reassess source classification and intrinsic properties.
Figure \ref{04-fig:ClassificationInitial} shows the confusion matrix for the initial candidates identified by the MBTA search on the \popone injection set.
We focus here on the $894$ injections for which an increase of \snr{} is observed between the initial trigger and the second iteration output.
When no such increase is observed, the released classification remains unchanged.
To construct this matrix, each injection is assigned its most probable classification following the method outlined in \cite{MBTAO4}. 
The results show that the MBTA pipeline tends to misclassify some sources. 
In particular, a significant fraction of NSBH injections are incorrectly identified as BBH, which may prevent triggering searches for counterparts.
Figure \ref{04-fig:ClassificationIt2} shows the confusion matrix for the candidates obtained by the second iteration, for which the classification is derived from the PE procedure detailed in Section \ref{sec:03}. 
We can see that NSBH sources are classified with significantly higher accuracy.
However, other source types appear to be less well identified: some BNS injections are misclassified as NSBH, and a few BBH as NSBH.
This behavior arises from the fact that the updated classification integrates over the full posterior distribution, while the standard MBTA classification relies on a single point-estimate. 
A significant portion of the posterior may lie in a different category than the injected one, which impacts classification.
Of the $53$ misclassified injections that were correctly classified after the initial detection, only $10$ had a probability below $10\%$ for their correct class. 
This mainly concerns BNS systems that are initially recovered by NSBH templates, due to an overdensity of high mass ratio templates in the search template bank.

\begin{figure}[!h]
    \centering
    \subfloat[HasNS]{\includegraphics[width=0.9\linewidth]{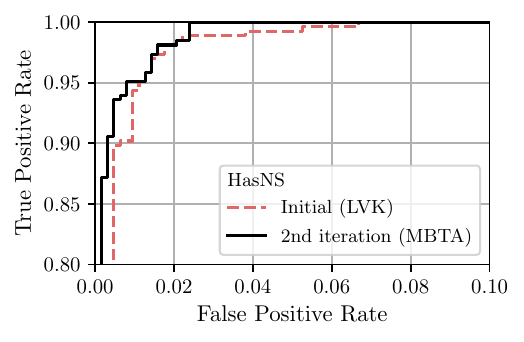}\label{04-fig:HasNS}} \\
    \subfloat[HasRemnant]{\includegraphics[width=0.9\linewidth]{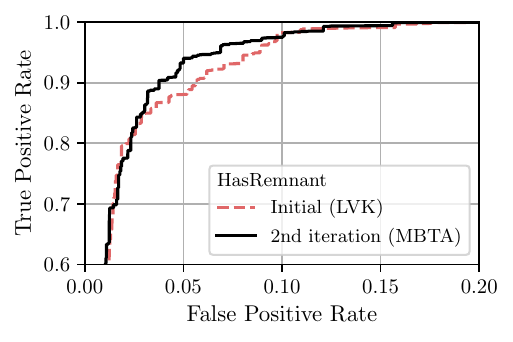}\label{04-fig:HasRemnant}} \\
    \subfloat[HasMassGap]{\includegraphics[width=0.9\linewidth]{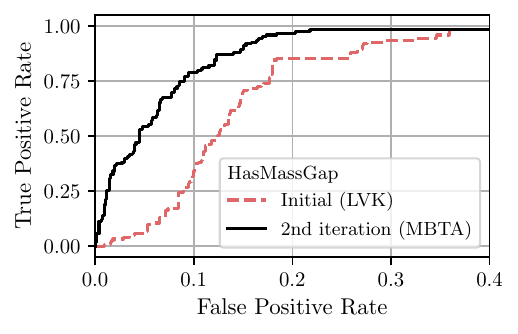}\label{04-fig:HasMassGap}}
    \caption{
    ROC curves of source properties made with a subset of O4a LVK common injections (\popone).
    The dotted red lines correspond to properties estimated from the initial MBTA triggers, using the algorithm outlined in \cite{Chaudhary_2024}.
    The black lines show results for the candidates produced by the second iteration of the \snropt{}, using the method from Section~\ref{sec:03}.
    }
    \label{04-fig:SourcePropRoc}
\end{figure}

Figure \ref{04-fig:SourcePropRoc} shows a comparison of receiver operating characteristic (ROC) curves for the source properties calculated before and after the \snropt{} procedure applied to the \popone injection set, considering only the $894$ injections for which we obtained an \snr{} gain.
The initial source properties are obtained with the standard LVK classifier, outlined in \cite{Chaudhary_2024}.
After applying  the \snropt{}, candidate properties are updated using the procedure described in Section \ref{sec:03}.
The results indicate that source properties are accurately recovered. 
In particular, the HasMassGap property is noticeably better identified compared to the standard LVK classifier.
Since this dataset does not include any SSM injections, we did not evaluate the performance of the corresponding classifier.
However, $31$ injections have a value of HasSSM above $10\%$.
All of these cases involve neutron stars and are initially recovered by highly asymmetric templates.

Releasing the \snropt{} classifications and properties would be appropriate even in the absence of an \snr{} gain, as they remain consistent with the results shown in Figures \ref{04-fig:ClassificationIt2} and \ref{04-fig:SourcePropRoc}, the PE procedure \tripledash{} based on posterior probability distribution integration \tripledash{} being robust to small \snr{} variations.

\section{The O4c implementation} \label{sec:05}
\subsection{Configuration differences}

Since June 13, 2025, the MBTA \snropt{} has been operating online.
Before September 5, we adopted a preliminary and simpler approach compared to what is presented in this paper:
\begin{itemize}
    \item All metrics for the search template bank were precomputed. 
    Consequently, the second iteration used the same metric as the first.
    \item No prior was applied to center the first iteration bank.
    Instead, the bank center and its associated metric were chosen to be those of the template that maximized the network \snr{} recovered by the search.
    \item No scaling factor (denoted by $\boldsymbol{S}$ in Equation~\ref{A01-eq:U_to_theta}) was applied.
    \item The first iteration of the universal bank contained half as many points and therefore covered a smaller parameter space.
    \item The second iteration bank, distributed according to the expected likelihood (see Equation~\ref{A01-eq:template_density_it2}), was generated with a fixed variance of $0.2$, corresponding to an \snr{} of approximately $3.5$.
    This conservative choice was made to avoid overly narrow banks.
\end{itemize}

\begin{figure}[!h]
    \centering
    \includegraphics[width=\columnwidth]{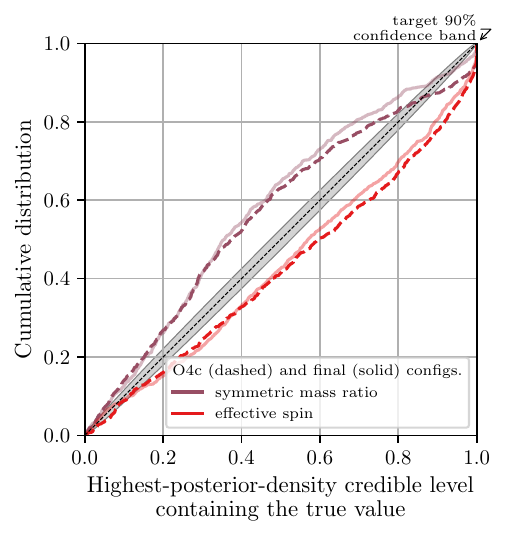}
    \caption{
    Comparison of P–P plots for mass ratio and effective spin recovered by the MBTA PE pipeline for a subset of O4a LVK common injections (\popone), using the online O4c configuration (dashed line) and the final configuration (solid line). 
    The black dashed line represents the diagonal, indicating unbiased posterior distributions with optimal width, while the grey area shows the $90\%$ confidence band derived from the beta distribution.
    }
    \label{05-fig:pp_plots_O4cVersion}
\end{figure}

The results in terms of \snr{} and sky map improvements were comparable to those obtained with the final configuration.
Using the \popone injection set, we measured a median \snr{} gain of $2.3\%$ and a median reduction of $3.4\%$ in the sky map searched area between the initial detection and the output of the second iteration with the O4c configuration.
Nevertheless, these changes lead to more accurate estimation of the source parameters.
As an illustration, Figure \ref{05-fig:pp_plots_O4cVersion} compares the P–P plots for symmetric mass ratio and effective spin obtained with the online O4c version applied to the \popone injection set, with those from the final configuration (already shown in Figure \ref{04-fig:pp_plots_pop1}).
The variances of the bias decrease by around $20\%$ and $10\%$ for the symmetric mass ratio and the effective spin, respectively. 
The average $90\%$ symmetric credible intervals (CI) for the effective spin are also reduced by about $8\%$, leading to a P–P plot closer to the diagonal.
For the symmetric mass ratio, the CI reduction is only about $4\%$ in average, which explains why the corresponding P–P plot deviates upward.
These improvements can be attributed to better centering and scaling of the banks.

\subsection{Online latency}

\begin{figure*}
    \centering
    \includegraphics[width=0.9\textwidth]{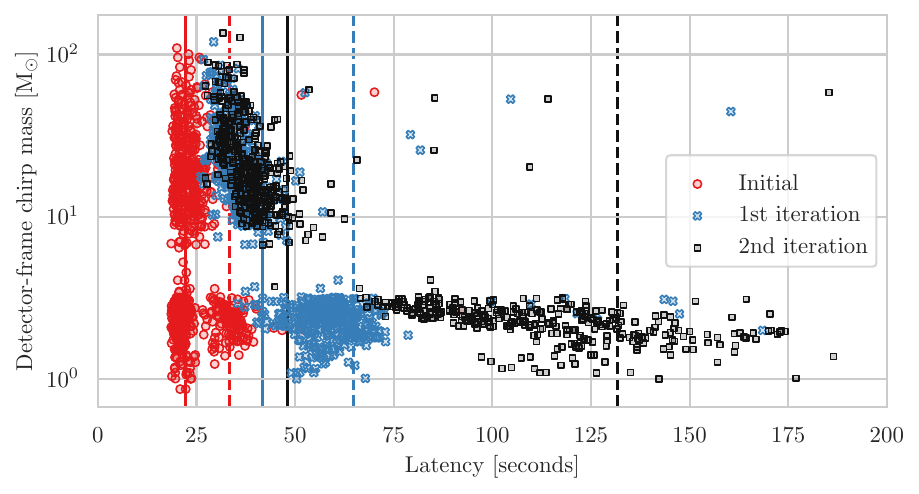}
    \caption{
    Upload latency from MBTA to GraceDB between June 13 and September 05, 2025.
    Solid (dashed) lines indicate the $50\%$ ($90\%$) quantiles of the latency distributions, at $22$, $42$, and $48$ seconds ($33$, $65$, and $132$ seconds) for the initial upload and the first and second \snropt{} iteration uploads, respectively.  
    Latencies are measured relative to the arrival time of the GW signal on Earth.
    The reported chirp mass corresponds to the highest-\snr{} template identified by each iteration.
    }
    \label{05-fig:LatencyUpload}
\end{figure*}

Both the \snropt{} and the PE procedure presented here are designed for low-latency operation, with results uploaded to GraceDB shortly after the pipeline identifies a potential CBC candidate.
Up to September 05, 2025, the MBTA search had reported 855 candidates (mostly of low significance) on which the procedure described here was applied.
These events were used to assess the latency of the \snropt{}.
Figure \ref{05-fig:LatencyUpload} shows the MBTA upload latency \tripledash{} defined as the time elapsed between the arrival of the GW signal on Earth and the submission of a candidate \tripledash{} plotted as a function of the detector-frame chirp mass of the highest-SNR template, for the O4c MBTA candidates submitted to GraceDB.
This latency does not include the additional time required to format and upload the MBTA PE results (including updated source classification, posterior samples, and monitoring plots), which was adding up to $1$ minute during O4c.
Most of these candidates are classified as noise transients and did not trigger public alerts.
We can see that the latency is strongly dependent on the masses.
This is mainly due to the time needed to generate low-mass templates with the \SEOBNRFOUROPT waveform model.
For low-mass systems, the first \snropt{} iteration takes about one minute after the initial detection to upload its results, with the second iteration adding approximately another minute.
In contrast, BBH signals (the most commonly detected sources) typically complete both iterations in under a minute after the arrival of the GW in the detector. 

\section{Conclusion} \label{sec:06}
We have presented a method implemented as a follow up to the MBTA pipeline that optimizes the \snr{} of CBC candidates and is designed for low-latency use.
The MBTA \snropt{} relies on an estimation of the metrics in the dimensionless chirp time space to predict how the \snr{} varies across the parameter space, allowing the placement of new templates in regions of interest.
By projecting generic sample banks on this space using the metrics, template placement can be performed in about one second.
The new templates are then used to filter the data using a configuration of MBTA optimized for handling templates of varying durations, with parallelized processing to reduce latency.
This procedure is applied in two iterations, following a hierarchical approach to mitigate potential biases in the parameters initially recovered by the search.

We have demonstrated that this technique statistically improves SNR recovery by approximately $2.5\%$, consistently with the expected coverage of the search template bank.
This increase in SNR translates into improved sky localization, with the search area typically reduced by $3.5\%$.

We also demonstrated how the results of the \snropt{} procedure can be used to quickly produce probability distributions of the source parameters.
This enables a more accurate source classification, particularly improving the identification of NSBH systems and potential EM-bright sources.
This approach is well suited to meet future challenges in CBC searches, where providing richer low-latency information will be essential to support follow-up efforts.
We show that the PE results obtained from the MBTA \snropt{} are in agreement with those recently reported by more extensive analyses of the LVK collaboration.

The MBTA \snropt{} has been running and uploading its results to GraceDB since June 13, 2025, providing updated \snr{} values and source classifications for the pipeline candidates.
This is achieved with an additional latency of approximately one to two minutes.
The time needed to update the triggers is mostly driven by the template waveform generation, in particular for low-mass systems.

\section*{Acknowledgments}  \label{sec:acknowledgments}
We express our gratitude to Gijs Nelemans and Jérôme Novak for their work in carrying out the internal LVK review of the MBTA pipeline and for their helpful comments and suggestions on the method presented in this paper.
We acknowledge Gregory Ashton for reviewing this paper as part of the LVK internal review process and for providing valuable feedback.
We thank our LVK collaborators from the CBC and low-latency groups for constructive comments.
This analysis exploits the resources of the computing facility at the EGO-Virgo site, and of the Computing Center of the Institut National de Physique Nucléaire et Physique des Particules (CC-IN2P3 / CNRS). 
This research has made use of data or software obtained from the Gravitational Wave Open Science Center (gwosc.org), a service of the LIGO Scientific Collaboration, the Virgo Collaboration, and KAGRA. 
This material is based upon work supported by NSF's LIGO Laboratory which is a major facility fully funded by the National Science Foundation, as well as the Science and Technology Facilities Council (STFC) of the United Kingdom, the Max-Planck-Society (MPS), and the State of Niedersachsen/Germany for support of the construction of Advanced LIGO and construction and operation of the GEO600 detector. 
Additional support for Advanced LIGO was provided by the Australian Research Council. 
Virgo is funded, through the European Gravitational Observatory (EGO), by the French Centre National de Recherche Scientifique (CNRS), the Italian Istituto Nazionale di Fisica Nucleare (INFN) and the Dutch Nikhef, with contributions by institutions from Belgium, Germany, Greece, Hungary, Ireland, Japan, Monaco, Poland, Portugal, Spain. 
KAGRA is supported by Ministry of Education, Culture, Sports, Science and Technology (MEXT), Japan Society for the Promotion of Science (JSPS) in Japan; National Research Foundation (NRF) and Ministry of Science and ICT (MSIT) in Korea; Academia Sinica (AS) and National Science and Technology Council (NSTC) in Taiwan.

\appendix

\section{Details on template placement} 
\label{sec:A01_universalbank}
\subsection{From ellipsoidal to spherical coordinates}
As discussed in Section \ref{sec:02}, the MBTA \snropt{} relies on a template bank which is obtained through projection of a universal set of samples.
By rearranging Equation \ref{02-eq:match_metric_approx}, the \textit{distance} $d_{AB}$ between two points $A$ and $B$ \tripledash{} whose coordinates are expressed in terms of dimensionless chirp times (defined in Equation \ref{02-eq:chirp_times}) \tripledash{} can be written as a function of the mismatch between the two corresponding waveforms trough a vector equation:
\begin{equation}
    d_{AB}^2 = 1 - M(\boldsymbol{\theta}_A, \boldsymbol{\theta}_B) = \Delta^{T} G \Delta,
    \label{A01-eq:distance_matrix}
\end{equation}
where $G$ is a matrix of coefficients $g_{\mu \nu}$ (see Equation \ref{02-eq:metric_def}), $\Delta \equiv \boldsymbol{\theta}_B - \boldsymbol{\theta}_A$ and $\Delta^T$ its transposed.
As $G$ is by definition a symmetric matrix, this equation can be written as:
\begin{equation}
\begin{split}
    d_{AB}^2 
    &= (P^{-1} \Delta)^T D (P^{-1} \Delta), 
\end{split}
\end{equation}
where $P D P^{-1} = G$, with $P$ an orthogonal matrix and $D$ a diagonal matrix.
We can therefore define a vector $\boldsymbol{U}$ such as:
\begin{equation}
    \boldsymbol{U} = \left( P^{-1} \Delta \right) \odot \left[ \operatorname{diag}(D) \right]^{1/2},
    \label{A01-eq:new_var}
\end{equation}
where $\odot$ denotes the element-wise (Hadamard) product, and $\operatorname{diag}(D)$ is the vector formed by the diagonal elements of the matrix $D$ (i.e., the eigenvalues of $G$).
With this new choice of coordinates, the distance between $A$ and $B$ is given by Euclid's formula:
\begin{equation}
    d_{AB}^2 = \boldsymbol{U}_\mu \boldsymbol{U}^\mu,
    \label{A01-eq:distance}
\end{equation}
and therefore any point $B$ having a constant match with $A$ lies on a sphere of radius $d_{AB}$.

\subsection{Universal bank generation}

Constructing the bank for the \snropt{} boils down to generating templates around the parameters initially recovered to populate a sphere in a space where distances are defined by Equation \ref{A01-eq:distance}.
To achieve this, we follow the method described in \cite{Marsaglia_1972}, drawing each point’s coordinates from a standard normal distribution and scaling the resulting vector to the desired radius.
Each new sample then represents a potential new template, covering a small region of the parameter space defined by a small exclusion sphere. 
The radius of this sphere is determined by evaluating Equation \ref{A01-eq:distance_matrix} at the desired local minimal match.
The sample is accepted into the bank only if no previously accepted point lies within its exclusion sphere.

As discussed at the end of Section \ref{sec:02}, we consider two different approaches.
In the first iteration of the \snropt{}, a constant minimal match of $98\%$ is used, resulting in templates uniformly distributed in dimensionless chirp times.
At this stage, $4000$ potential templates are retained, represented in the top row of Figure \ref{02-fig:bank_conversion_example}.
In the second iteration, templates are distributed to focus around the most probable point identified by the first pass.
We adopt a sampling strategy designed to match the template distribution to the likelihood expressed in Equation \ref{03-eq:likelihood}. 
Templates are placed such that the probability density of the match $M$ with respect to the bank center, characterized by $\boldsymbol{\theta}_O$, is given by:
\begin{equation}
    f_T(\boldsymbol{\theta}) = f_0 \exp\!\left[-\tilde{\rho}^2 \left(1 - M(\boldsymbol{\theta}_O, \boldsymbol{\theta})\right)\right],
    \label{A01-eq:template_density_it2}
\end{equation}
where $\tilde{\rho}$ is defined as
\begin{equation}
    \tilde{\rho} = \tilde{\rho}_{\max} \frac{k \bar{\rho}}{k \bar{\rho} + \tilde{\rho}_{\max}},
    \label{A01-eq:template_density_it2_SNR}
\end{equation}
with $\bar{\rho}$ denoting the mean SNR measured across detectors for the template yielding the highest network \snr{}, $k$ a safety factor (typically set to $0.4$) used to artificially broaden the bank, and $\tilde{\rho}_{\max}$ a parameter meant to set an upper limit on $\tilde{\rho}$, fixed to $15$ to ensure that the second iteration bank encompasses more than one template from the first iteration.
The normalization constant $f_0$ is chosen so that the resulting bank contains $7000$ potential templates.

\subsection{Sample projection}

The second and third steps of the MBTA \snropt{} involve projecting the universal template bank, discussed above, into the physical parameters used to generate waveforms. 
This requires inverting Equation \ref{A01-eq:new_var} to recover the chirp times of each potential template:
\begin{equation} 
    \boldsymbol{\theta}_T = \boldsymbol{\theta}_O + P_O \left( \boldsymbol{S} \odot \boldsymbol{U}_T \odot \left[\operatorname{diag}(D_O) \right]^{-1/2} \right),
    \label{A01-eq:U_to_theta}
\end{equation}
where the vectors $\boldsymbol{U}_T$ and $\boldsymbol{\theta}_T$ represent the template $T$ in the universal and dimensionless chirp-time spaces, respectively.
The matrices $P_O$ and $D_O$ contain the eigenvectors and eigenvalues, respectively, of the metric evaluated at the bank center $\boldsymbol{\theta}_O$.
The vector $\boldsymbol{S}$ is used to scale each dimension independently and is typically set to $(1,1,2)$ for the first iteration and $(1,1,1)$ for the second iteration.

Next, the dimensionless chirp times are converted back into component masses and spins using Equation \ref{02-eq:chirp_times_to_phys}. 
To resolve the missing spin degree of freedom, we equally distribute the reduced spin between the two components.
Any chirp-time combinations that do not correspond to physically valid masses and spins \tripledash{} $\mtot^d < 0$, $\eta \notin (0, 0.25]$ or $|\spineff| > 1$ \tripledash{} are discarded.
Templates with $\mtot^d > 500$ or $m_1/m_2 > 100$ are also excluded.
The complete projection procedure is illustrated in Figure \ref{02-fig:bank_conversion_example}.

\section{Prior calculation}
\label{sec:A02_priorcalculation}
In standard LVK parameter inference, an agnostic prior is typically assumed. 
This prior is taken to be uniform within some plausible range in individual detector-frame component masses $(m_1^d, m_2^d)$ and spin magnitudes $(\chi_1, \allowbreak \chi_2)$, while isotropic in spin orientations \cite{GWTC4_methods}. 
However, the parameter estimation technique presented here relies on samples expressed in terms of the dimensionless chirp times $\boldsymbol{\theta} \equiv (\theta_0, \theta_3, \theta_{3s})$, defined in Equation \ref{02-eq:chirp_times}.
Since individual spins are constrained to be identical by the bank generation process, the spin dependency in $\boldsymbol{\theta}$ is encapsulated in a single spin component:
\begin{equation}
    \chi^z = \chi_1 \cos(\phi_1) = \chi_2 \cos(\phi_2),
    \label{A02-eq:spin_z}
\end{equation}
with $(\phi_1, \phi_2)$ the spin orientation angles relative to the orbital angular momentum of the binary.
Therefore the prior PDF used in Equation \ref{03-eq:bayes} can be expressed as:
\begin{equation}
    f(T(\boldsymbol{\lambda} \equiv \boldsymbol{\theta})) = |J| \cdot f_{m_1^d, m_2^d, \chi^z}(m_1^d, m_2^d, \chi^z)
    \label{A02-eq:prior_calculation}
\end{equation}
where $|J|$ denotes the absolute value of the determinant of the Jacobian matrix:
\begin{equation}
    \begin{split}
        |J| 
        &= 
        \begin{vmatrix}
            \frac{\partial m_1}{\partial \theta_0} & \frac{\partial m_2}{\partial \theta_0} & \frac{\partial \chi^z}{\partial \theta_0} \\[4pt]
            \frac{\partial m_1}{\partial \theta_3} & \frac{\partial m_2}{\partial \theta_3} & \frac{\partial \chi^z}{\partial \theta_3} \\[4pt]
            \frac{\partial m_1}{\partial \theta_{3s}} & \frac{\partial m_2}{\partial \theta_{3s}} & \frac{\partial \chi^z}{\partial \theta_{3s}}
        \end{vmatrix}
        \\
        &= \left| \frac{75 \sqrt[3]{10} \omega_{\eta}^{-2} c^{6} G^{-2} \Msun^{-2}}{1216 \sqrt[3]{10} \pi \theta_{0}^{3} - 4520 \pi^{-\frac{2}{3}} \theta_{0}^{\frac{7}{3}} \theta_{3}^{\frac{5}{3}}} \right|,
        \label{A02-eq:jacobian}
    \end{split}
    \end{equation}
with $\omega_{\eta}^2 = 4 \pi^{2} f_{0}^{2} \sqrt{1 -4 \eta}$.
Assuming a standard prior where masses and spins are uncorrelated and masses are uniformly distributed, the prior PDF simplifies: 
\begin{equation} 
    \begin{split} 
        f_{m_1^d, m_2^d, \chi^z}(m_1^d, m_2^d, \chi^z) \propto f_{\chi^z}(\chi^z). 
    \end{split} 
    \label{A02-eq:joint_mass_spin_proba}
\end{equation}

The aligned-spin $\chi^z$ PDF corresponds to the distribution of the product of two uncorrelated, uniformly distributed variables, the spin magnitude and its relative projection along the orbital angular momentum:
\begin{equation} 
    f_{\chi^z}(\chi^z) = -\frac{1}{2} \log_e\left(|\chi^z|\right).
    \label{A02-eq:aligned_spin_proba}
\end{equation} 

Combining Equation \ref{A02-eq:jacobian} and Equation \ref{A02-eq:aligned_spin_proba} gives the prior PDF expressed in dimensionless chirp times. 
To prevent the prior from diverging at $\eta = 0.25$ (Equation \ref{A02-eq:jacobian}) or $\chi^z = 0$ (Equation \ref{A02-eq:aligned_spin_proba}), the parameters used in its evaluation are clipped to $q = m_2 / m_1 \leq 0.99$ and $|\chi^z| \geq 0.001$.
Reweighting by the PDF used for template placement yields the prior probability distribution, illustrated in Figure \ref{03-fig:prior_example}.

\section{Detector-frame into source-frame posterior conversion} 
\label{sec:A03_frameconversion}
The parameter estimation method presented in Section \ref{sec:03} enables inference of the binary’s detector-frame masses and spins.
To correct for redshift magnification and obtain a distribution of source-frame parameters, we use the marginal luminosity distance probability distribution provided by the Bayestar algorithm \cite{Singer_2016}, applied to the highest-SNR trigger.
The MBTA \snropt{} evaluates the joint probability distribution $p_{m_1^d, m_2^d, \spineff}$ at several thousand points across the parameter space.
To compensate for the limited number of samples, the conversion problem is simplified by assuming that only the mass ratio is correlated with the spin, resulting in a 2D frame-dependent individual mass probability distribution:
\begin{equation}
    \begin{split}
        &p_{m_1^d, m_2^d, \spineff}(m_1^d, m_2^d, \spineff) \\
        &\approx p_{m_1^d, m_2^d}(m_1^d, m_2^d) \cdot p_{\spineff | \eta}(\spineff | \eta).  
    \end{split}
    \label{A03-eq:mass_spin_split}
\end{equation}

Source-frame masses ($m_1^s, m_2^s$) are related to detector-frame masses through the source redshift $z$:
\begin{equation}
    m^s 
    = \frac{m^d}{1 + z}.
    \label{A03-eq:source_frame_def}
\end{equation}

A change of variables enables expressing the source-frame joint mass and redshift probability distribution in terms of the detector-frame probability distribution:
\begin{equation}
    \begin{split}
        &p_{m_1^s, m_2^s, z} (m_1^s, m_2^s, z) \\
        &= (1+z)^2 p_{m_1^d, m_2^d, z} \left(m_1^s (1+z), m_2^s (1+z), z \right),
    \end{split}
    \label{A03-eq:source_frame_change_of_var}
\end{equation}
Since our analysis marginalizes over extrinsic parameters, distance is not directly measured.
We therefore rely on the luminosity distance probability distribution $p_{D_L}$ estimated by Bayestar, based on the highest-\snr{} trigger. 
To obtain an estimate of the redshift probability distribution, we again apply a change of variables:  
\begin{equation}
    p_{z}(z) = \frac{\partial f_{z \rightarrow d_L}^{-1}(d_L)}{\partial d_L} p_{d_L}\left(d_L = f_{z \rightarrow d_L}(z)\right),
    \label{A03-eq:redshift_pdf}
\end{equation}
where the relation $d_L = f_{z \rightarrow d_L}(z)$ is derived from a standard cosmological model, as described in \cite{Planck2015}. 
The distance probability distribution produced by Bayestar relies on the assumption that the mass parameters are that of the highest-\snr{} trigger, and would be different for different masses, due to the dependence of the GW inspiral amplitude (primarily) on the chirp mass \cite{Chatziioannou_2024}. 
To take this into account, we define the conditional redshift probability for given source-frame component masses as a homothetic transform of the probability distribution in Equation~\ref{A03-eq:redshift_pdf}:
\begin{equation}
    p_{z}(z \mid m_1^s, m_2^s) = p_{z}\left(z_{\mathrm{eff}}\right),
    \label{A03-eq:redshift_pdf_eff}
\end{equation}
where $z_{\mathrm{eff}}$ is the effective redshift corresponding to a rescaled luminosity distance:
\begin{equation}
    d_{L,\mathrm{eff}} = f_{z \rightarrow d_L}(z) \cdot \left(\frac{m_{\mathrm{chirp,highest\text{-}SNR}}^d}{\mchirp^s (1+z)}\right)^{5/3},
\end{equation}
with $m_{\mathrm{chirp,highest\text{-}SNR}}^d$ the detector-frame chirp mass of the highest-\snr{} trigger (for which the redshift probability is known) and $\mchirp^s$ the source-frame chirp mass for which we want to estimate the redshift probability value.

As a result, the joint distribution of Equation \ref{A03-eq:source_frame_change_of_var} can be factorized and marginalized over redshift to get the source-frame individual mass probability distribution:
\begin{equation}
    \begin{split}
        &p_{m_1^s, m_2^s} (m_1^s, m_2^s) \\
        & = \int p_{m_1^s, m_2^s, z} (m_1^s, m_2^s, z) dz \\
        & = \int p_{m_1^d, m_2^d}\big((1+z)m_1^s, (1+z)m_2^s\big) \\
        & \hspace{2em} (1+z)^2 p_z(z \mid m_1^s, m_2^s) dz.
        \end{split} 
    \label{A03-eq:joint_mass_redshift_pdf}
\end{equation}
To provide a usable data product, this probability distribution is sampled using rejection sampling.
A comparison of the detector-frame and source-frame posterior probability distributions obtained with this method is shown in Figure \ref{03-fig:post_GW230529}.

\bibliographystyle{style}
\bibliography{bibliography.bib} 

\end{document}